\documentclass[preprint,12pt,authoryear]{elsarticle}

\usepackage{amssymb}

\journal{Chaos, Solitons and Fractals}

\begin{document}

\begin{frontmatter}



\title{Spreading of infections on random graphs: A percolation-type model for COVID-19}


\author{Fabrizio Croccolo\footnote{Corresponding author}}
\address{Universite de Pau et des Pays de l'Adour, E2S UPPA, CNRS, TOTAL, LFCR UMR5150, 
             Anglet, France. (email: fabrizio.croccolo@univ-pau.fr)}

\author{H. Eduardo Roman}
\address{Department of Physics, University of Milano-Bicocca, 
              Piazza delle Scienze 3, 20126 Milan, Italy. (email: eduardo.roman@mib.infn.it)}

\begin{abstract}
We introduce an epidemic spreading model on a network using concepts from percolation theory. The model
is motivated by discussing the standard SIR model, with extensions to describe effects of lockdowns within
a population. The underlying ideas and behavior of the lattice model, implemented using the same lockdown 
scheme as for the SIR scheme, are discussed in detail and illustrated with extensive simulations. A comparison 
between both models is presented for the case of COVID-19 data from the USA. Both fits to the empirical data 
are very good, but some differences emerge between the two approaches which indicate the usefulness of 
having an alternative approach to the widespread SIR model.
\end{abstract}




\end{frontmatter}

\section{Introduction}
\label{sect:Intro}
The study of epidemics spreading in human populations has a long history both on the mathematical aspects (see e.g.
\cite{kermack1927contribution,bailey1957mathematical,bollobas2013modern}), as well as on the modelling of outbreak and control 
of their evolution (\cite{anderson757oxford,fraser2004factors,Villaverde2020}). Spreading phenomena has been studied extensively also in the realm of statistical physics of disordered systems (e.g. \cite{ben2000diffusion,bunde2012fractals,stauffer2018introduction}).

In this paper, we introduce a lattice network model for epidemics sprea\-ding based in part on concepts taken from percolation theory. 
To motivate the network approach to spreading, we first discuss the SIR model (\cite{kermack1927contribution}), also extending it to encompass the effects of lockdowns mimicked by using a time decaying reproduction number. To assess the usefulness of the lattice model, we consider COVID-19 data from the USA and compare the results of the simulations with SIR predictions, both in the presence 
of lockdowns.

The paper is organized as follows: We start out in Sect.~\ref{sect:sirmodels} with a brief review of the SIR model, with emphasis on 
some analytical results and its extension to the description of lockdown effects. Illustrative examples are shown, together with a motivation for the need of going beyond a `mean-field' approach. The network model is then discussed in detail in Sect.~\ref{sect:Network}, and the percolation ideas, relevant to the present case, are discussed. Extensive (Monte Carlo) simulations are shown to illustrate the advantages and difficulties typically associated with such lattice models. In Sect.~\ref{sect:covid19}, we apply it to the current case of COVID-19 USA data, by comparing the network results with the predictions of the SIR model. The paper ends with our concluding remarks in Sect.~\ref{sect:conclusions}.  

\section{SIR models with extensions to lockdown effects}
\label{sect:sirmodels}

Infection spreading is typically modelled by the SIR model, introduced by \cite{kermack1927contribution}. We briefly review it for the purpose of introducing the notation, presenting extensions to describe lockdown effects. We follow standard terminology in epidemic literature\footnote{It is widespread usage in epidemiology (see e.g. \cite{smith2001sir}) to refer to `Susceptibles', `Infecteds' and `Recovereds', rather than using longer phrases such as `population of susceptible individuals' or `the susceptible category'. Here, we add the `Dormants' category, referring to individuals who momentarily do not interact
with others.}.

\begin{figure}[!h]
\begin{center}
\includegraphics[width=7cm, height=7cm]{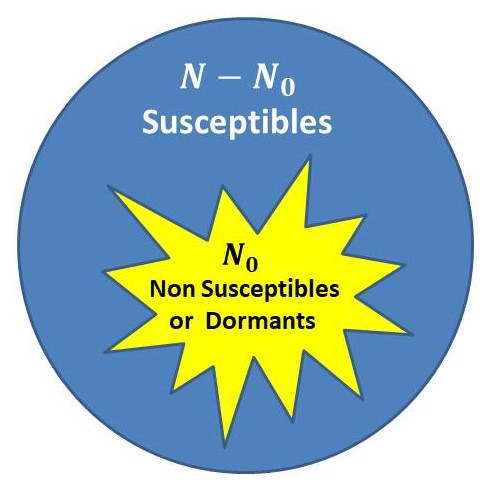}
\caption{Structure of a population of $N$ individuals before the outbreak of the disease: Susceptibles, $N-N_0$, 
             versus non-susceptibles or dormants, $N_0$. The latter are assumed to be inaccessible to the infection and   
             being disseminated uniformly within the population. Although this classification is apparently superfluous 
             within a SIR approach, it becomes useful for spreading phenomena on networks.}
\label{fig:fzero}
\end{center}
\end{figure}

We consider a population with a fixed number of individuals, $N$. To describe the outbreak of an infectious disease,
the population can be divided into the following four categories: Susceptibles ($S$), Infecteds ($I$), Recovereds ($R$)
and Dormants ($D$). The number of individuals in the first three categories depends on time $t$, so that we will indicate
them as $S(t)$, $I(t)$ and $R(t)$, while the number of dormants $D$ remains constant during the whole process. Susceptibles are initially healthy but can get infected; infecteds carry the infection and can transmit it to susceptibles, while recovereds are infecteds who have healed or have died, thus they do not spread the infection any further. The fourth category, $D$, includes non susceptibles (perhaps immune individuals to the specific disease), and in general healthy subjects who do not get in contact with infecteds during the whole spreading process (Fig.~\ref{fig:fzero}). The reason for introducing the fourth category will be clearer when discussing solutions of the SIR system of equations. 

At any time $t$, the number of individuals in each category must obey the conservation equation,
\begin{equation}             
S(t)+I(t)+R(t)+N_0 =N,
\label{eq:totalN}
\end{equation}
where we have denoted $N_0=D$. We assume the unit of time to be one day. 

The SIR model is described by a set of three differential equations for the time variation of the number of
individuals in each category, $S(t)$, $I(t)$ and $R(t)$, which, upon taking into account the condition Eq.~(\ref{eq:totalN}), read,
\begin{eqnarray}             
\frac{dS}{dt} &=& -\beta \, S(t) \frac{I}{N} = \bar\beta \, S(t) \frac{I}{N_{\rm eff}},      \\
\frac{dI}{dt}  &=&  \beta \, I(t) \frac{S}{N} - \gamma \, I(t) = \bar\beta \, I(t) \frac{S}{N_{\rm eff}} - \gamma \, I(t),      \\
\frac{dR}{dt} &=& \gamma \, I(t),    
\label{eq:SIRequationsNeff}
\end{eqnarray}             
where $N_{\rm eff}=N-N_0=S(t)+I(t)+R(t)$, is the effective number of indivi\-duals taking part in the process, $\beta$ is the infection  
(or contact) rate between infecteds and susceptibles, $\bar\beta=\beta N_{\rm eff}/N$ is the effective infection rate in the presence of dormants, and $\gamma$ is the healing (or immunization) rate of infecteds. It is convenient to work with normalized quantities, 
$s(t)=S(t)/N_{\rm eff}$, $i(t)=I(t)/N_{\rm eff}$, and $r(t)=R(t)/N_{\rm eff}$, so that the SIR equations become,
\begin{eqnarray}             
\frac{ds}{dt} &=& -\bar\beta \, s(t) i(t),                           \label{eq:SIRNeffnorm1}\\
\frac{di}{dt} &=&   (\bar\beta \, s(t) - \gamma) \, i(t) = (\bar R_0 s(t)-1) \gamma\, i(t),       \label{eq:SIRNeffnorm2}\\
\frac{dr}{dt} &=& \gamma \, i(t).                                    \label{eq:SIRNeffnorm3}
\end{eqnarray}             
The above equations have exactly the same form as in the case $N_0=0$. The idea of considering  explicitely a fraction of the whole population not taking part in the spreading phenomenon, $f_0=N_0/N$ ($0\le f_0\le 1$), allows us to interpret the so-called reproduction number, $R_0$, as composed of two factors, a purely `biological' one, $\sim \beta/\gamma$, and a `structural' one $\sim (1-f_0)$, denoted as, $\bar R_0=(\beta/\gamma) (1-f_0)$. The second factor represents the effect of dormants non-in-contact with others, thus `hindering' or slowing down the spreading process. The fraction $f_0$ can change in time, but for simplicity we assume it constant. Within the realm of the SIR model dormants don't seem necessary, however, they play a prominent role within the context of network models of infection spreading, as we will discuss in detail in Sect.~\ref{sect:Network}. More generally, dormants can be considered as those individuals who interact very weakly with others, thus representing a subset of the population which is in a sort of `quarantine' from others. 

One can derive some general relations between the categories by conside\-ring ratios between the SIR differential equations (Eqs.~(\ref{eq:SIRNeffnorm1},\ref{eq:SIRNeffnorm2},\ref{eq:SIRNeffnorm3})). First, divide (\ref{eq:SIRNeffnorm2}) by (\ref{eq:SIRNeffnorm1}), yielding,
\begin{equation}             
\frac{di}{ds}=-1+\frac{\gamma}{\bar\beta S}, \quad i(t)-i(0)=s(0)-s(t)+\frac{1}{\bar R_0}\log(s(t)/s(0)),
\label{eq:Ratio65}
\end{equation}
and dividing (\ref{eq:SIRNeffnorm1}) by (\ref{eq:SIRNeffnorm3}), we find,
\begin{equation}             
\frac{ds}{dr}=-\frac{\bar\beta S}{\gamma}, \quad \log(s(t)/s(0))=-{\bar R_0} (r(t)-r(0)).
\label{eq:Ratio57}
\end{equation}
By specifying the conditions, $i(0)=1/N_{\rm eff}$, $s(0)=1-1/N_{\rm eff}$, and $i(\infty)=0$, Eq.~(\ref{eq:Ratio65}) becomes,
\begin{equation}             
s(\infty)= 1 + \frac{1}{\bar R_0}\log(s(\infty)/s(0)),
\end{equation}
which can be written as,
\begin{equation}             
s(\infty)= s(0)\, e^{-\bar R_0 (1-s(\infty))}.
\label{eq:Ratio65Conditions}
\end{equation}
Notable limits are, $s(\infty)=s(0)$ when $\bar R_0\to0$, and $s(\infty)=0$ when $\bar R_0\to\infty$. Also Eq.~(\ref{eq:SIRNeffnorm2})
admits a partial solution when $di/dt=0$, in particular at $t=t_{\rm peak}$, i.e. at the peak of the infecteds curve, yielding,
\begin{equation}             
s(t_{\rm peak})= \frac{1}{\bar R_0}, \quad {\rm and} \quad i(t_{\rm peak})+r(t_{\rm peak})=1-\frac{1}{\bar R_0}.
\label{eq:Peakatispeak}
\end{equation}

\begin{figure}[!h]
\begin{center}
\includegraphics[width=10cm, height=7cm]{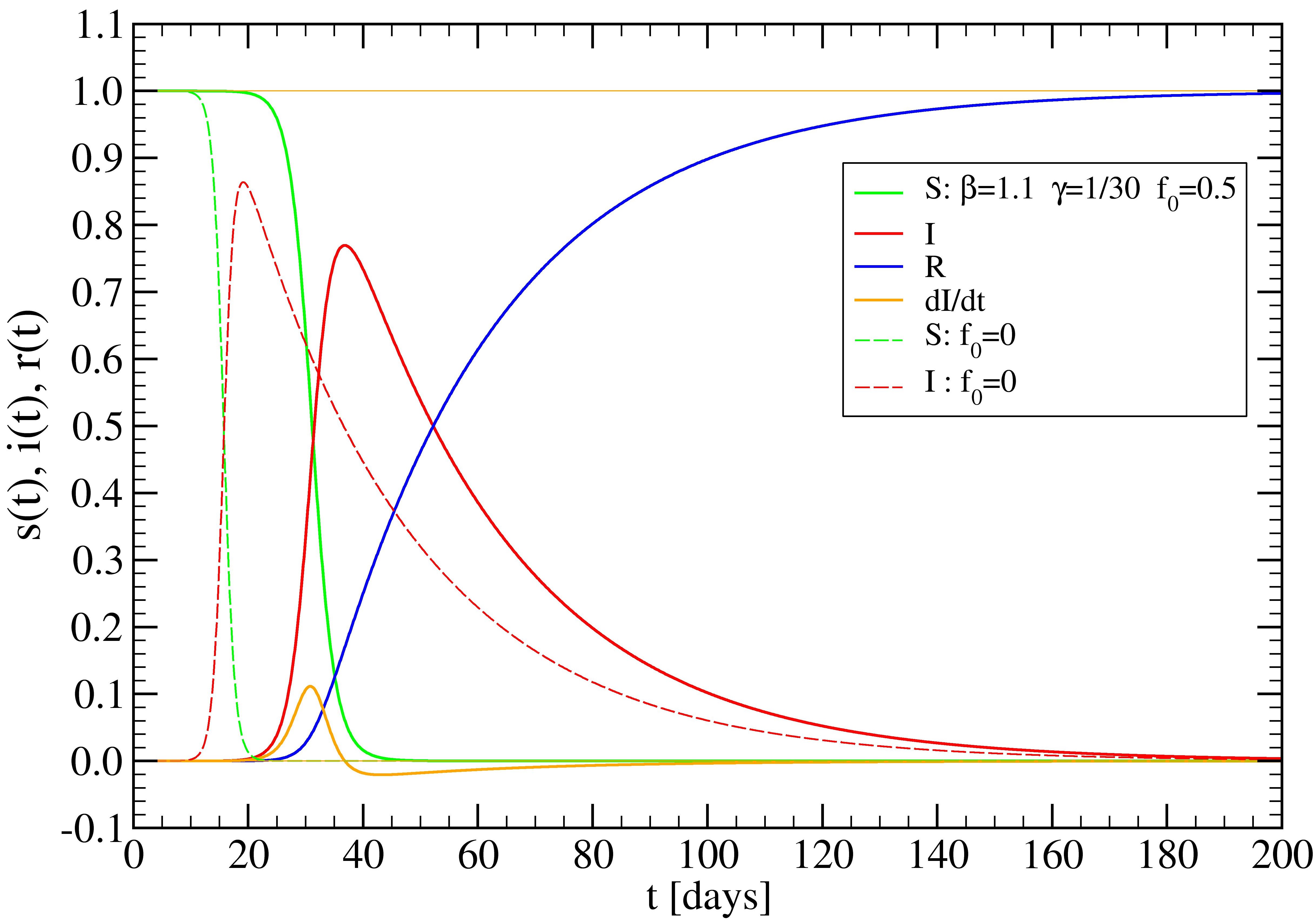}
\includegraphics[width=10cm, height=7cm]{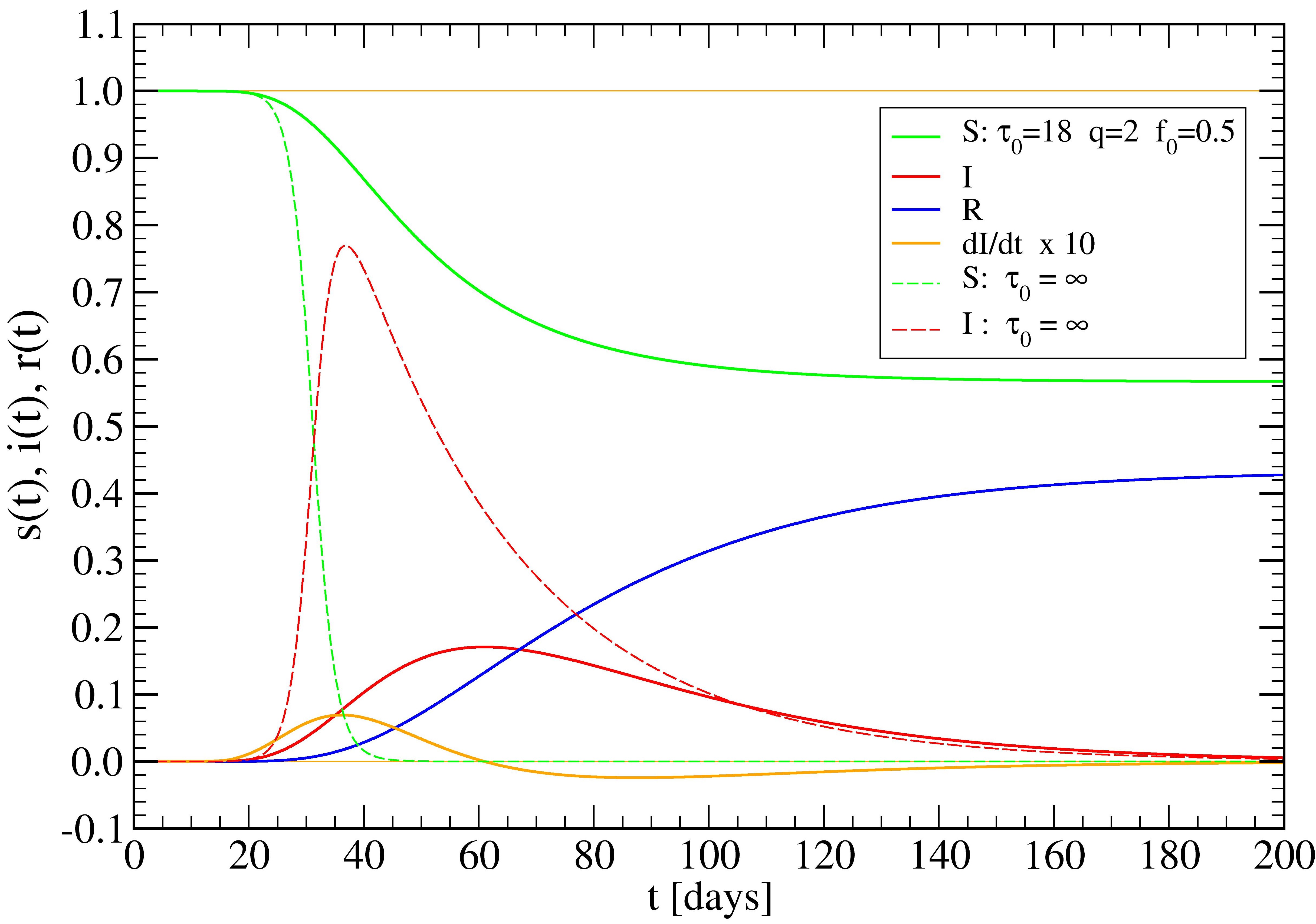}
\caption{(Upper panel) SIR model for $\beta=1.1$, $\gamma=1/30$, $R_0=\beta/\gamma=33$ and $f_0=N_0/N=0.5$,  
             yielding $\bar\beta=0.55$ and $\bar R_0=16.5$. The case $f_0=N_0=0$ (dashed lines) is shown for comparison.
             (Lower panel) SIR model with distancing effects for $\bar\beta(t)=\bar\beta (\tau_0/t)^q$, with $\tau_0=18$   
             and $q=2$. The dashed lines represent the solution without lockdowns, and are shown for comparison.}
\label{fig:sirnumericsD}
\end{center}
\end{figure}

Typically, lockdown effects are modelled using an exponential time dependence of $\beta$ (see e.g. \cite{Palladino2020}). 
Here, we employ a softer decay which appears to work very well, i.e.
\begin{equation}             
\bar\beta(t) = \bar\beta \left(\frac{\tau_0}{t}\right)^q, \, t\ge \tau_0,  
\label{eq:barbetatime}
\end{equation}
and $\bar\beta(t) = \bar\beta$, for $t\le\tau_0$, where $\tau_0$ is the time at which lockdowns start, and $q>0$ is a parameter. This
means we deal with a time decaying reproduction number, $\bar R_0(t)=\bar\beta(t)/\gamma$. Using this form in Eq.~(\ref{eq:SIRNeffnorm2}), we can obtain the time $t_{\rm lock}$ at which the infecteds curve displays its new maximum. The condition is, $\bar R_0(t_{\rm lock}) s(t_{\rm lock})=1$, which, together with Eq.~(\ref{eq:Peakatispeak}), yields,
\begin{equation}             
t_{\rm lock}= \tau_0 \left(\frac{s(t_{\rm lock})}{s(t_{\rm peak})}\right)^{1/q}> \tau_0, 
\label{eq:tpeaklockdowns}
\end{equation}
since $s(t_{\rm lock})>s(t_{\rm peak})$, as there are more susceptibles in the presence of lockdowns than otherwise. Illustrative 
examples are reported in Fig.~(\ref{fig:sirnumericsD}), for $\bar R_0=16.5$ (with $\beta=1.1$ and $\gamma=1/30$) and $f_0=1/2$.

\newpage

In the upper panel, we show the standard case, with $\gamma$ representing the inverse of a typical COVID-19 healing period. We
show also for comparison the case $f_0=0$ (dashed lines). In this case, Eq.~(\ref{eq:Ratio65Conditions}) predicts $s(\infty)\cong 0$,
clearly consistent with the numerical data. As is apparent from the upper panel of Fig.~(\ref{fig:sirnumericsD}), the effect of
$f_0>0$ is to shift the infecteds peak at longer times by keeping a still high reproduction number. The effects of lockdowns, using
relation Eq.~(\ref{eq:barbetatime}), are displayed in the lower panel of the figure, for the typical cases $\tau_0=18$~days and $q=2$.
Notice that according to Eq.~(\ref{eq:tpeaklockdowns}), $t_{\rm lock}\cong 18 \sqrt{0.66/0.06}\cong 60$~days, and the infecteds
peak is reduced by a factor of about 4.

Let us return to Eq.~(\ref{eq:Ratio65Conditions}). In the language of random graph theory 
(\cite{erdHos1960evolution,erdos1973art,bollobas2013modern}), Eq.~(\ref{eq:Ratio65Conditions}) is formally equiva\-lent
to the self-consistent relation for the fraction of nodes, $P_\infty\equiv 1-s(\infty)$, belonging to the giant component of random graphs with a finite mean node degree $\left<k\right>$, given by,
\begin{equation}             
P(\infty)= \left(1-e^{-\left<k\right>p\,P(\infty)}\right),
\label{eq:fractionGiantcluster}
\end{equation}
where $0<p<1$ is the probability of occupancy of a link between two nodes. It can be shown that a giant cluster exists when
$\left<k\right>p=c>1$. This model is also related to the mean-field theory of random spin glasses with finite coordination number
(see e.g. \cite{kanter1987mean}). Clearly, $P(\infty)\to 1$ if $c\to\infty$, while for $c-1=\varepsilon$, with $0\le\varepsilon\ll1$ we find,
\begin{equation}             
P(\infty) = 2p \frac{c-1}{c^2}, \quad {\rm when} \quad c\to 1^+.
\label{eq:fractionGiantclustercto1}
\end{equation}
The correspondence with Eq.~(\ref{eq:Ratio65Conditions}) is achieved if we take $\bar R_0\equiv c$ and assume $s(0)=1$, which is the
case since $s(0)\approx 1$ in our spreading model. Using $1-s(\infty)=r(\infty)$, we find,
\begin{equation}             
r(\infty) = 2 \frac{\bar R_0-1}{\bar R_0^2}, \quad {\rm when} \quad \bar R_0\to 1^+.
\label{eq:fractionGiantclustercto1}
\end{equation}
This correspondence is actually not surprising since the SIR equations are valid in a mean-field sense, where fluctuations and correlations 
among the categories are neglected. This analogy suggests us that we should go beyond mean-field theory by studying infectious spreading on a network where correlations can be implemented. This is done in the following Section.

\section{Spreading phenomena on random graphs: Percolation concepts}
\label{sect:Network}

Tracing infecteds in a population and how they move is essential to make an accurate assessment of the extent of which a virus has spread in a region, country or the whole world, in order to implement effective lockdowns in each particular place (see e.g. \cite{Vespignani2020}). Here we discuss a simple network model defined on a two dimensional square lattice. The sites of the lattice represent individuals belonging to one of the four categories ($S$, $I$, $R$, $D$), which we will distinguish with different colors in the plots, i.e. green, red, blue, and yellow, respectively. 

Two individuals are said to be connected, i.e. transmission of the disease can occur, if they are nearest-neighbors (NN) on the lattice, representing a `short-range' contact interaction. The NN choice is done just for convenience, and it can be relaxed in other versions of the model. The bonds between sites represent therefore the links in the graph, and the coordination number of 4 gives the maximum
node degree, under `static' conditions. The latter mean that the individuals are considered to be at rest in their lattice locations, initially, while the virus can move around from site to site if the following rules are obeyed: (1) The virus can cross a bond from an infected site to a susceptible one; (2) no virus transmission occurs otherwise; (3) infected sites heal after $\tau_H$ days, becoming recovered sites, so that they can neither infect others nor being infected again (immune sites); and (4) dormant sites do not participate in the spreading process.

It is essential to consider additional links `dynamically' as the spreading goes on. This is done in order to describe those individuals
who move around for different reasons. Thus, any infected site can reach sites which are not NN to it, and the infection can spread according to the above rules. In this case, the infection is transmitted with a probability 
$\beta_{\rm L}= 1/\tau_{\rm L}$, with $\tau_{\rm L}> 1$, provided the target site is a susceptible one.

In summary, we have taken a two dimensional lattice to facilitate the visualization of the network, and considering both NN transmissions
as well as long-range ones, though with a lower probability. Since the extra links are not determined from the beginning, but are added dynamically, we do not show them in the plots, facilitating the identification of the categories. Another reason for choosing a square lattice is that the percolation threshold for site percolation (in this case the susceptible sites) is about 0.6, meaning that if we take say, $f_0=1/2$, there is no percolating (`infinite') cluster of susceptibles on the lattice. This `hindering' effect is useful when implemen\-ting lockdowns, since the remaining susceptible clusters (of connected NN sites) are disconnected from each other (they are indeed `finite').
One can say that `long-range' links can connect different susceptible clusters, which otherwise would remain disconnected.

\begin{figure}[!h]
\begin{center}
\includegraphics[width=6.5cm, height=6.5cm]{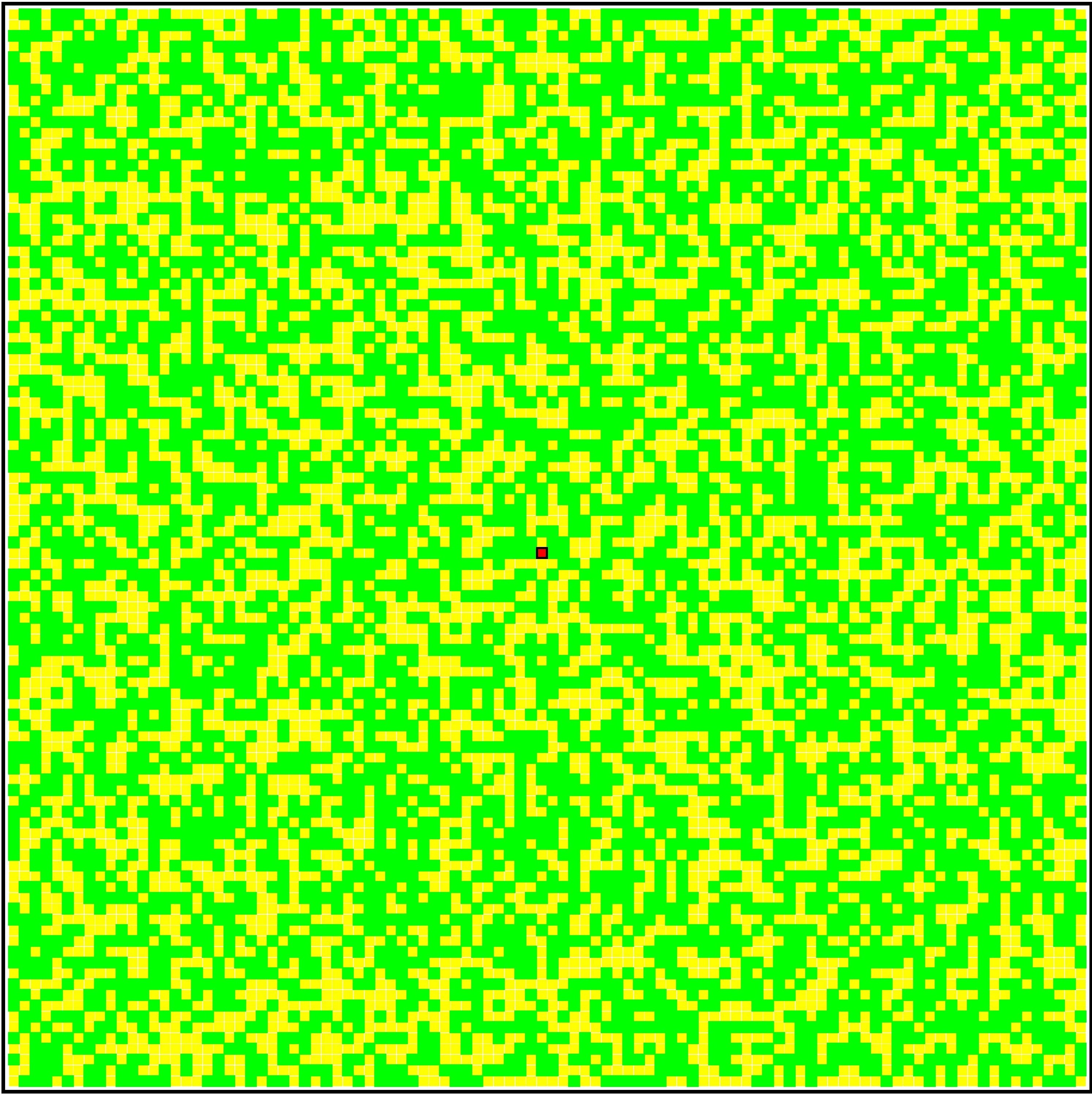}
\includegraphics[width=6.5cm, height=6.5cm]{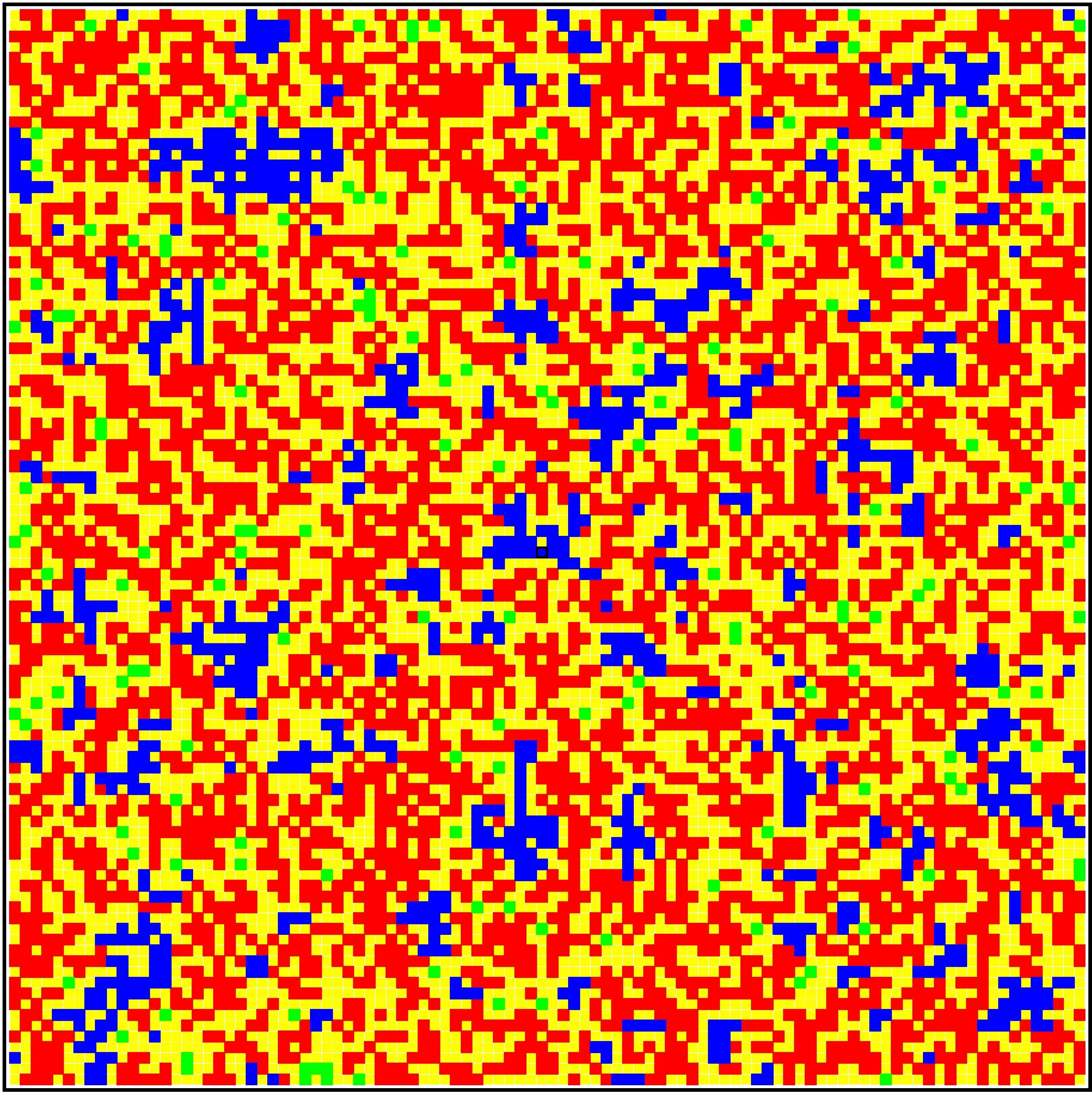}
\includegraphics[width=6.5cm, height=6.5cm]{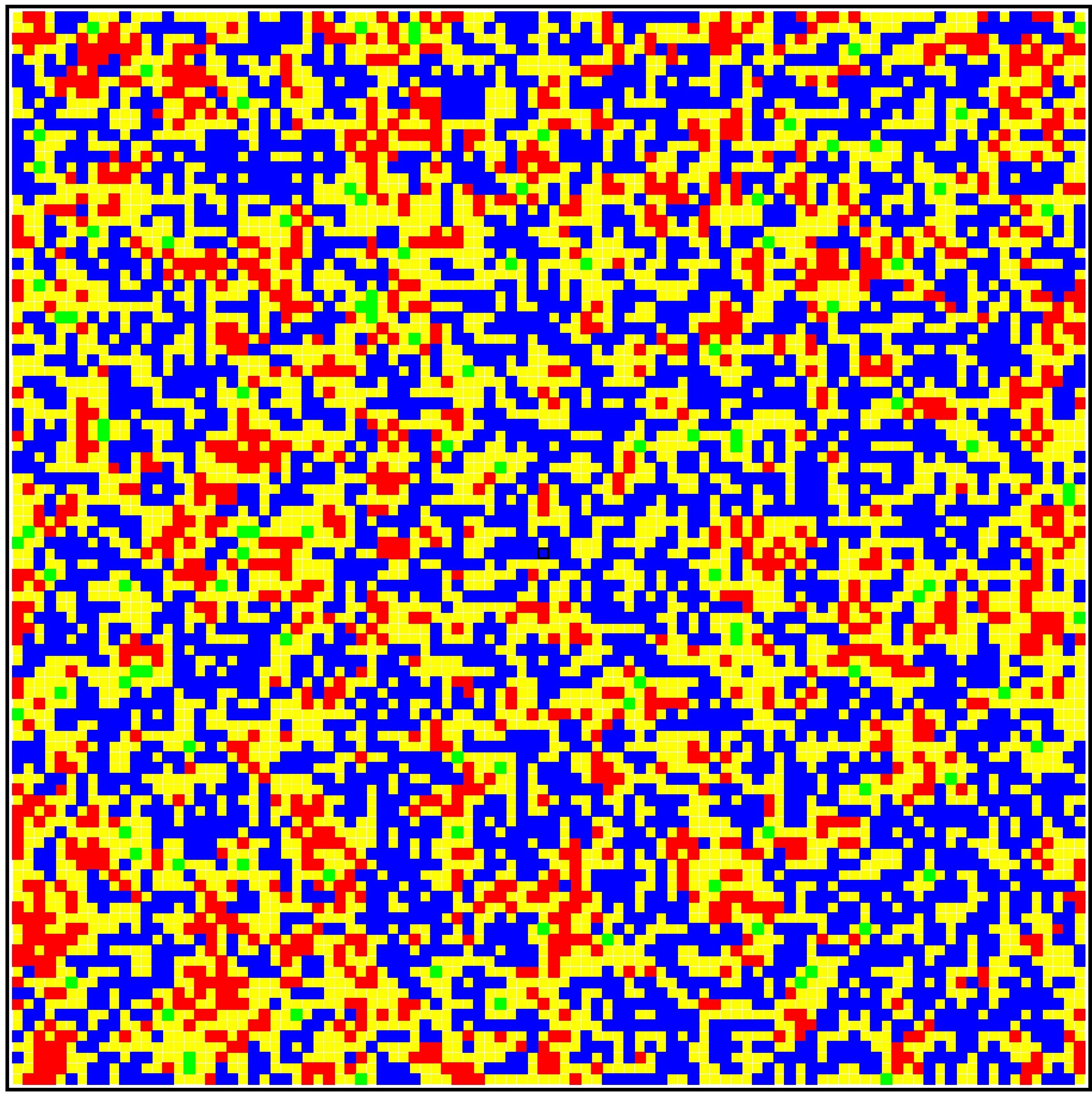}
\includegraphics[width=6.5cm, height=6.5cm]{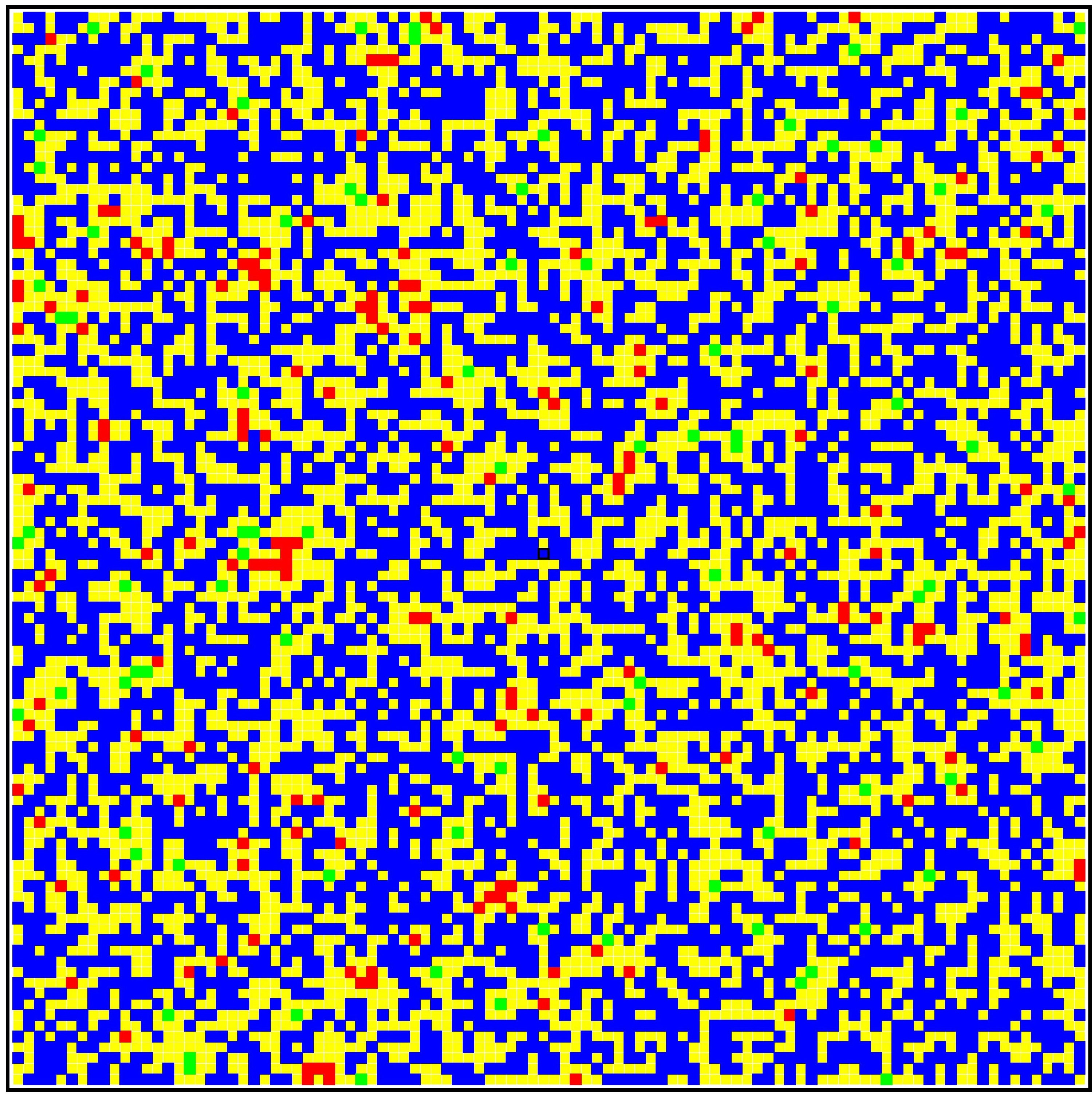}
\caption{The graph of connected individuals used in the simulation. Each site of the (100x100) square lattice represents an individual
             belonging to one of the four categories: (S) Susceptible (green), (I) Infected (red), (R) Recovered (blue), (D) Dormant 
             (yellow). {\bf Panels}: (Upper left) Starting configuration ($t=0$) for $f_0=0.5$ with $S=5042$, $I=1$, $R=0$ and $D=4957$; 
             (Upper right) $t=50$; (Lower left) $t=60$; (Lower right) $t=70$. The model parameters are: $\tau_{\rm I}=2$ (transmission 
             time) and $\beta=1/\tau_I=1/2$, $\tau_{\rm H}=30$ (healing time) and $\gamma=1/30$, yielding $R_0=\beta/\gamma=15$, 
             and $\tau_{\rm L}=10$ (long range transmision time) yielding $\beta_{\rm L}=1/10$. Times are expressed in days. The 
             average node degree for the starting configuration is $\left<k\right>=2$, while additional links are added dynamically as the 
             network evolves in time. The newly created links yield an additional mean degree 
             $\left<\Delta k\right>=1110/5043\cong0.22$, corresponding to an effective mean node degree 
             $\left<k_{\rm eff}\right>=2.22$.}
\label{fig:ConfigNetwork}
\end{center}
\end{figure}

\begin{figure}[!h]
\begin{center}
\includegraphics[width=6.5cm, height=6.5cm]{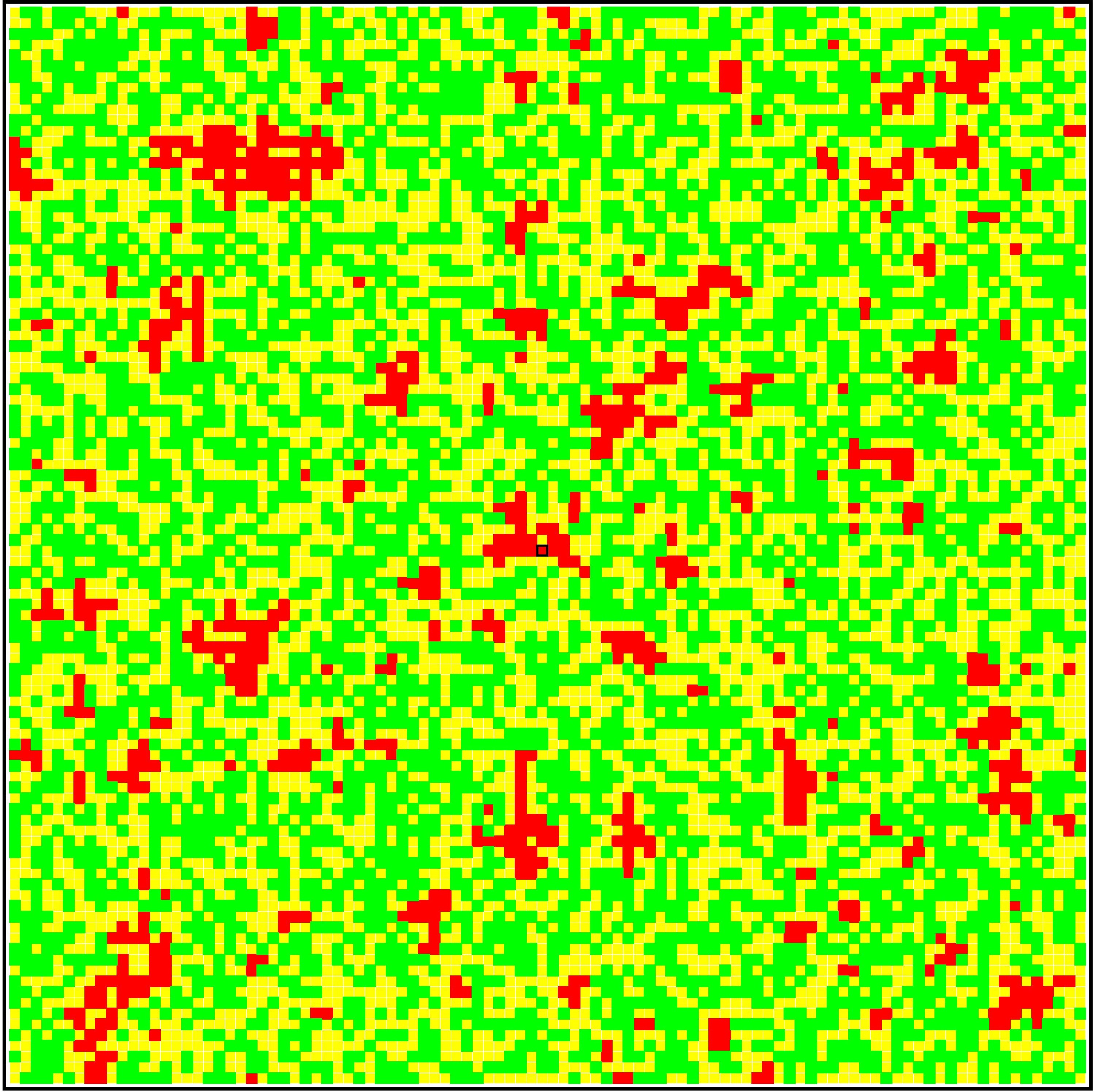}
\includegraphics[width=6.5cm, height=6.5cm]{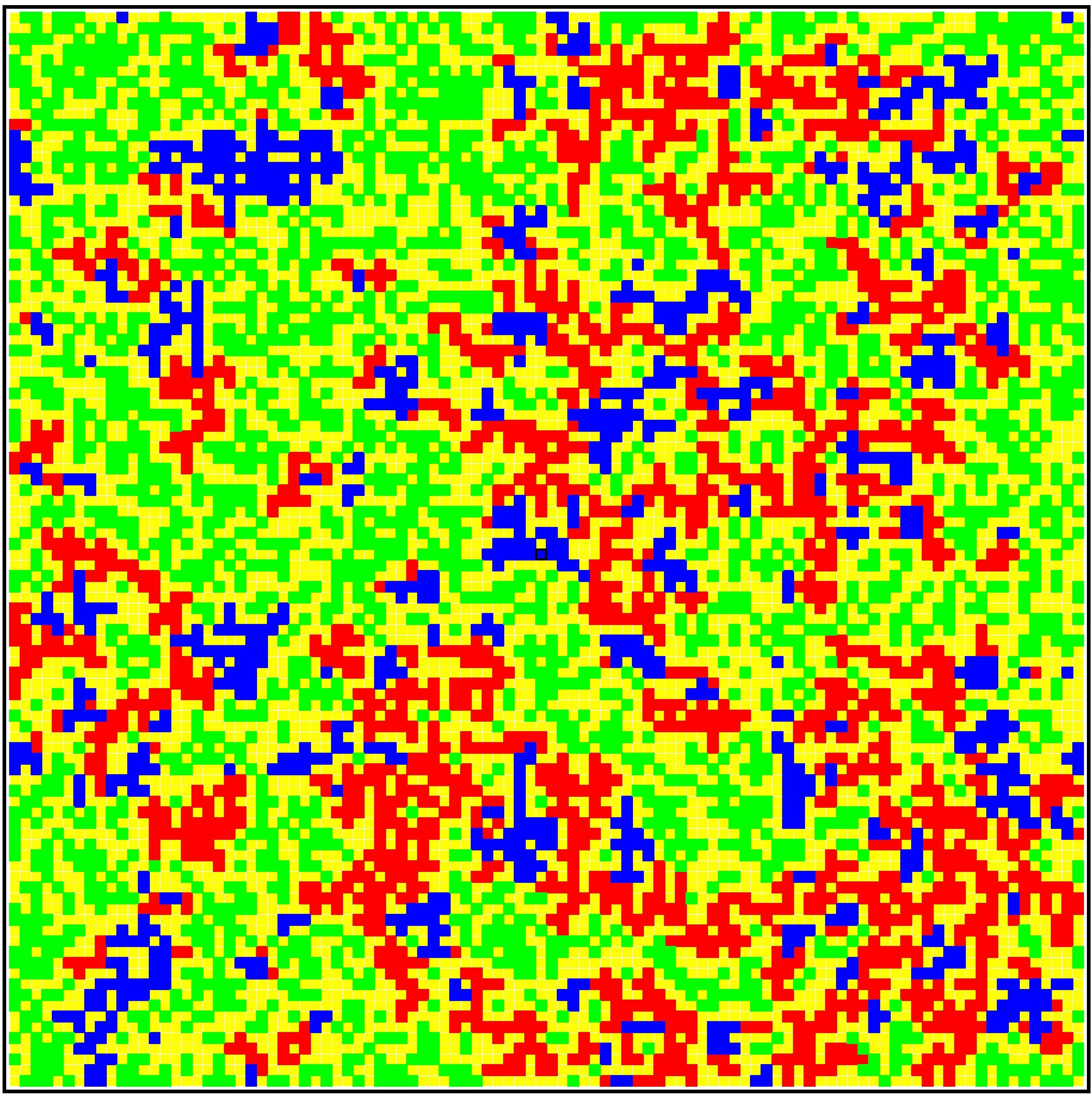}
\includegraphics[width=6.5cm, height=6.5cm]{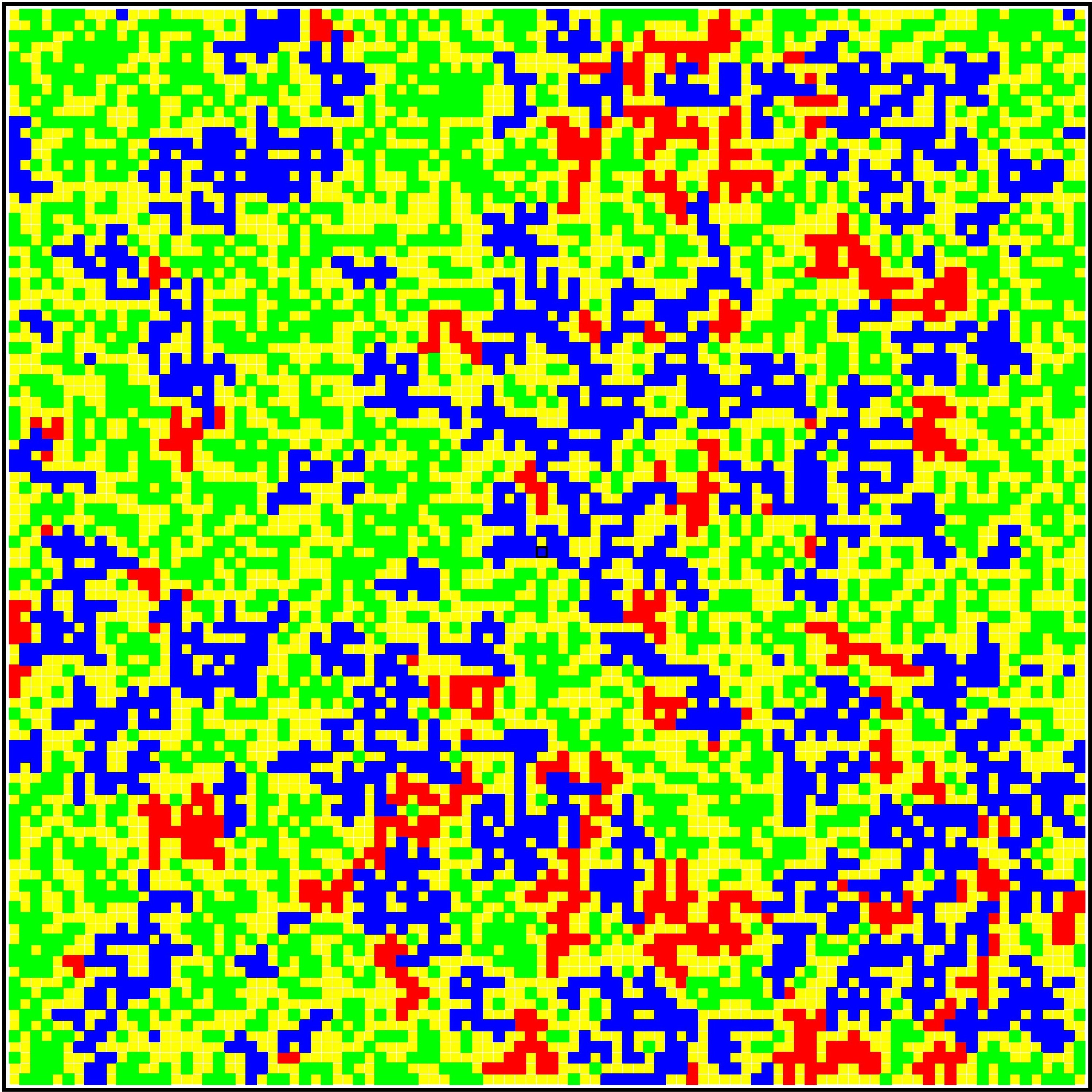}
\includegraphics[width=6.5cm, height=6.5cm]{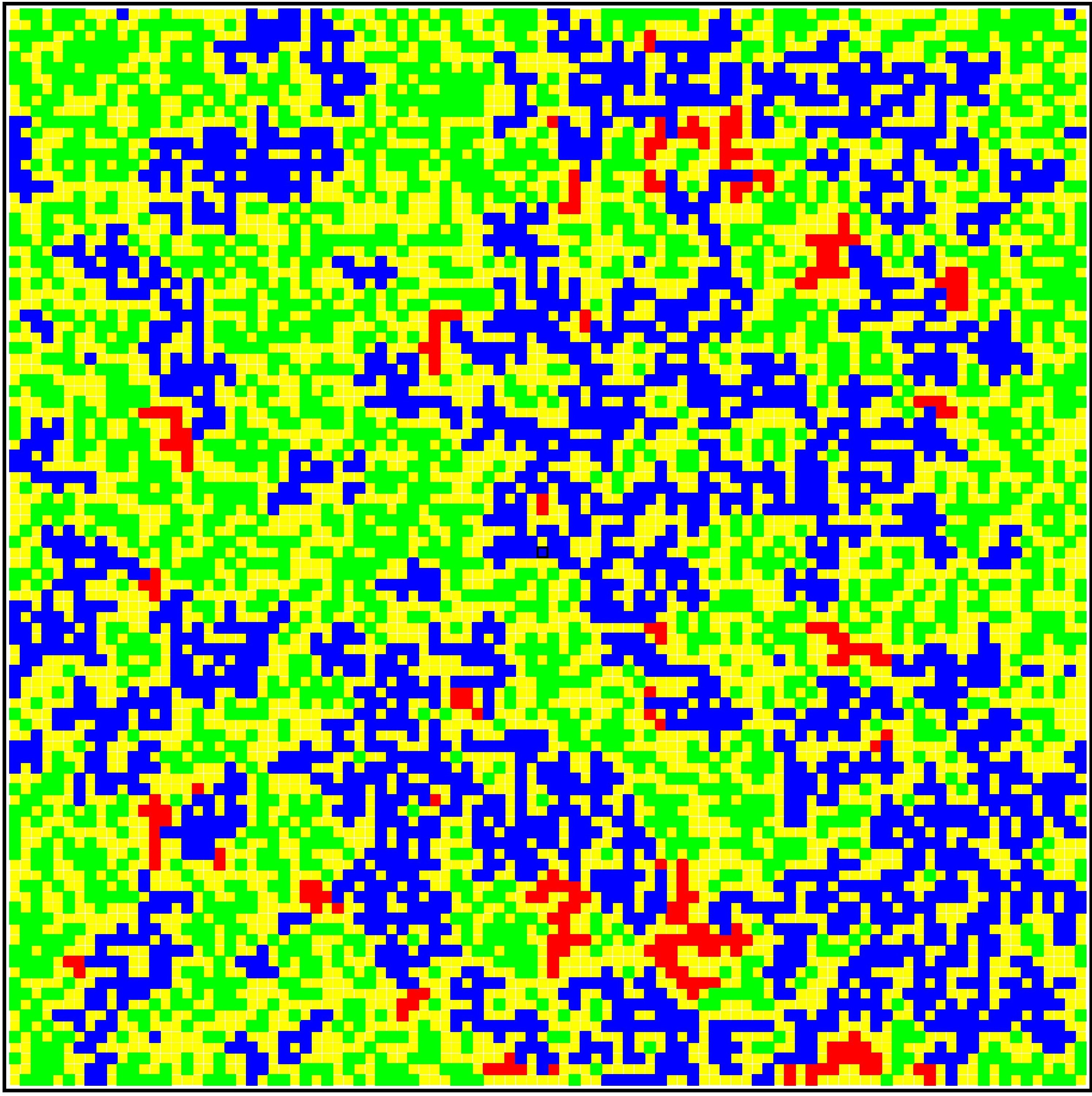}
\caption{Same as in Fig.~\ref{fig:ConfigNetwork} in the case of lockdowns: {\bf Panels}: (Upper left) Starting lockdowns ($t=20$); 
             (Upper right) $t=50$; (Lower left) $t=60$; (Lower right) $t=70$. The model parameters are: $\tau_0=20$ (start of lockdowns) 
             and $\tau_{\rm L}=\infty$ (no long range transmission time) yielding $\beta_{\rm L}=0$. The average node degree for the 
             starting configuration is $\left<k\right>=2$, while additional links are added dynamically until $t=\tau_0$. The newly 
             created links yield an additional mean degree $\left<\Delta k\right>=156/5043\cong0.03$, corresponding to an effective mean 
             node degree $\left<k_{\rm eff}\right>=2.03$.}
\label{fig:ConfigNetworkD}
\end{center}
\end{figure}

We show in Fig.~\ref{fig:ConfigNetwork} and \ref{fig:ConfigNetworkD} results of simulations for a single configuration, 
without and with lockdowns, respectively, on a (100x100) lattice for times $t\le 100$~days. Fixed boundary conditions 
are employed. The starting configuration has a random distribution of either susceptible or dormant sites, chosen with probability 
$f_0=1/2$. The initial conditions include a single infected site right at the center of the lattice, and none recovereds. 
We keep track of the existing infected sites, each carrying a clock which starts ticking when the site gets infected. After a 
time $\tau_{\rm L}$ it becomes recovered (immune). Death sites are not implemented, but they can be estimated simply as a 
fraction of recovereds.

\begin{figure}[!h]
\begin{center}
\includegraphics[width=9cm, height=6.5cm]{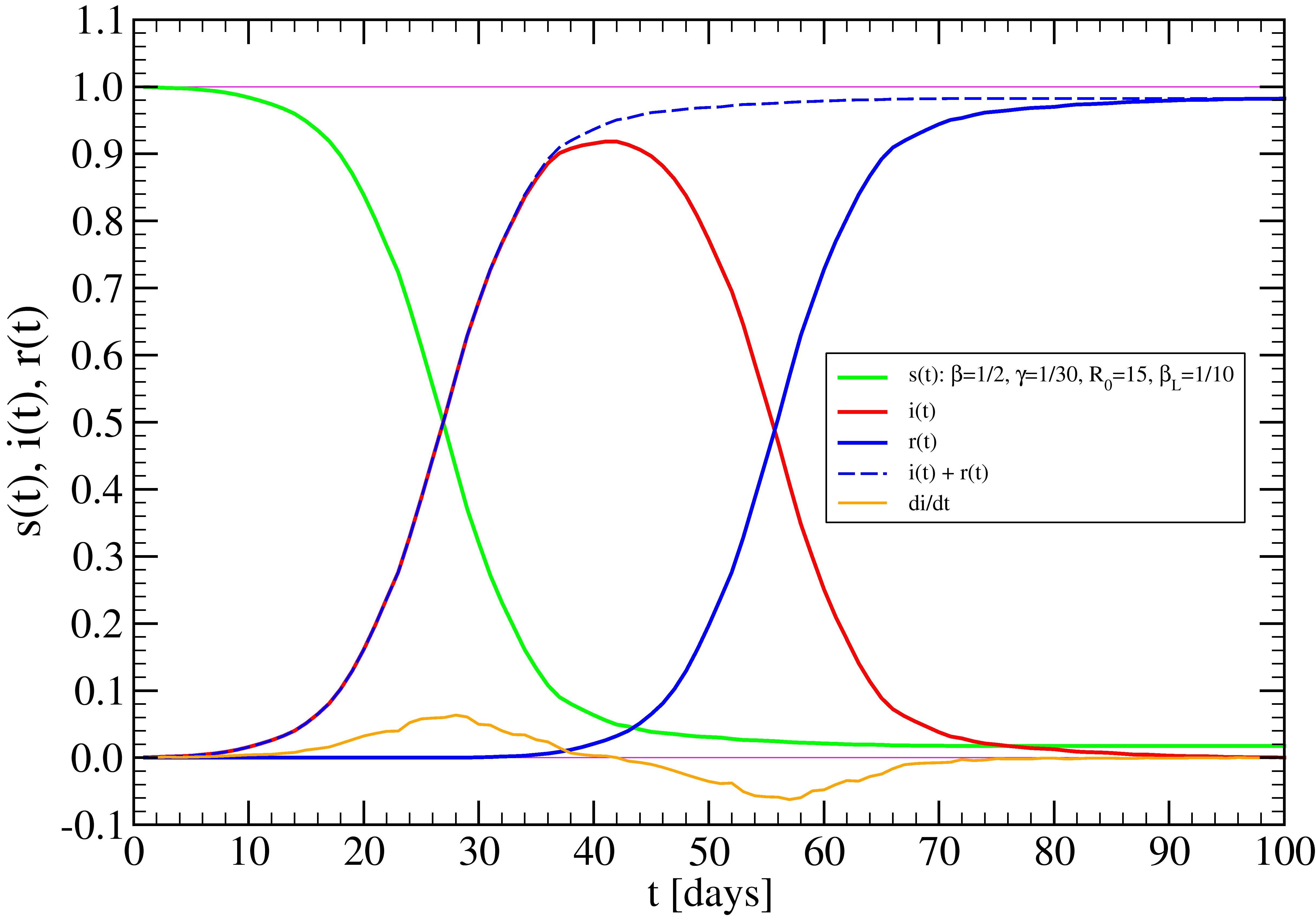}
\includegraphics[width=9cm, height=6.5cm]{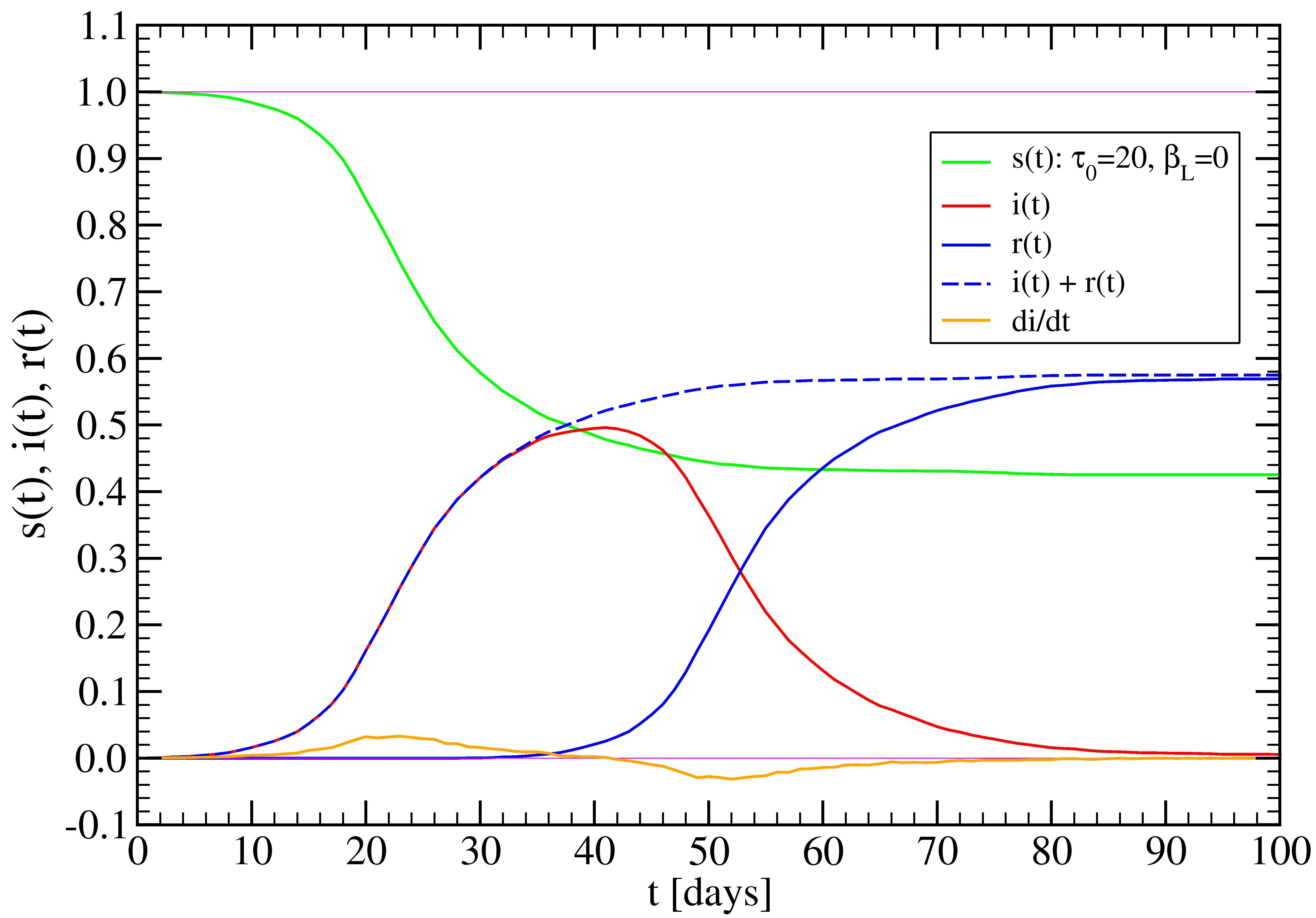}
\caption{Time evolution of the spreading model of Fig.~\ref{fig:ConfigNetwork}. (Upper panel) No lockdowns.
             (Lower panel) Lockdowns: $t\ge \tau_0=20$.}
\label{fig:timeevolutionNetwork}
\end{center}
\end{figure}

We count the number of dynamical links generated during the spreading, from which we can determine, a posteriori, the effective mean node degree, $\left<k_{\rm eff}\right>$, in our network. We find $\left<k_{\rm eff}\right>\cong 2.22$ without lockdowns 
(Fig.~\ref{fig:ConfigNetwork}) and $\left<k_{\rm eff}\right>\cong 2.03$ with lockdowns (Fig.~\ref{fig:ConfigNetworkD}), 
indicating that effectively, $\left<k_{\rm eff}\right>f_0\gtrsim 1$. It turns out that the relatively small reduction of the mean node 
degree in the presence of lockdowns is sufficient to reduce the number of infecteds considerably, as one can see from the very
different structure of recovereds clusters from both figures. The time evolution of the three categories are displayed in 
Fig.~\ref{fig:timeevolutionNetwork}, and look qualitatively similar to those from the SIR model in Fig.~\ref{fig:sirnumericsD}.
We should mention that a single lattice simulation takes few seconds on a typical laptop, even for lattices of size (400x400), allowing to obtain accurate mean values by averaging over several configurations.

\section{Analisys of COVID-19 USA data}
\label{sect:covid19}
As an application of the present ideas we consider COVID-19 USA data (see also \cite{wu2020estimating,dong2020interactive,xu2020pathological}), from the point of view of both SIR and network models. 
In order to do so, and due to the complexity of the data, we need to introduce additional features in particular for the SIR
model. 

\begin{figure}[!h]
\begin{center}
\includegraphics[width=10.5cm, height=7.45cm]{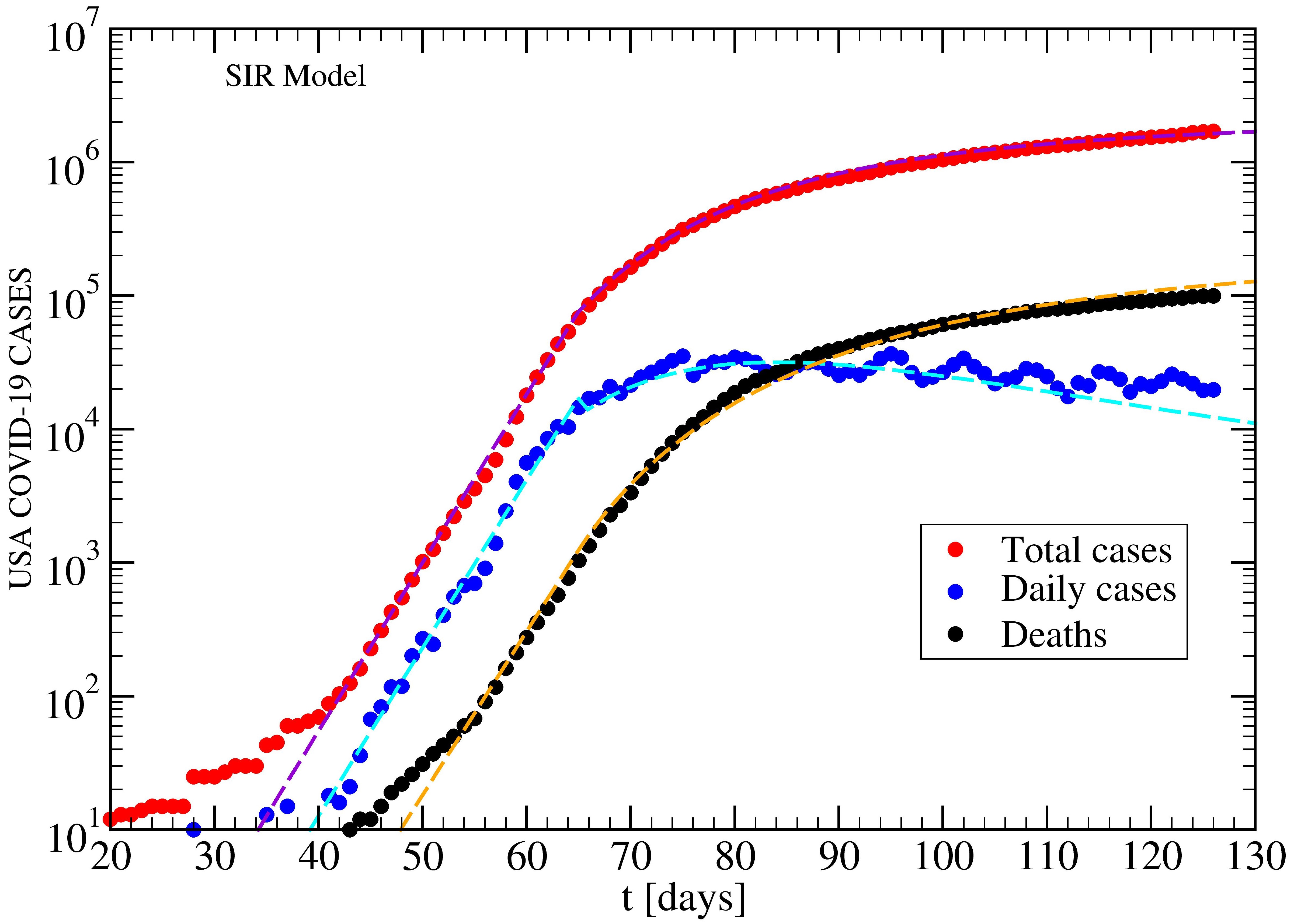}
\includegraphics[width=10.5cm, height=7.45cm]{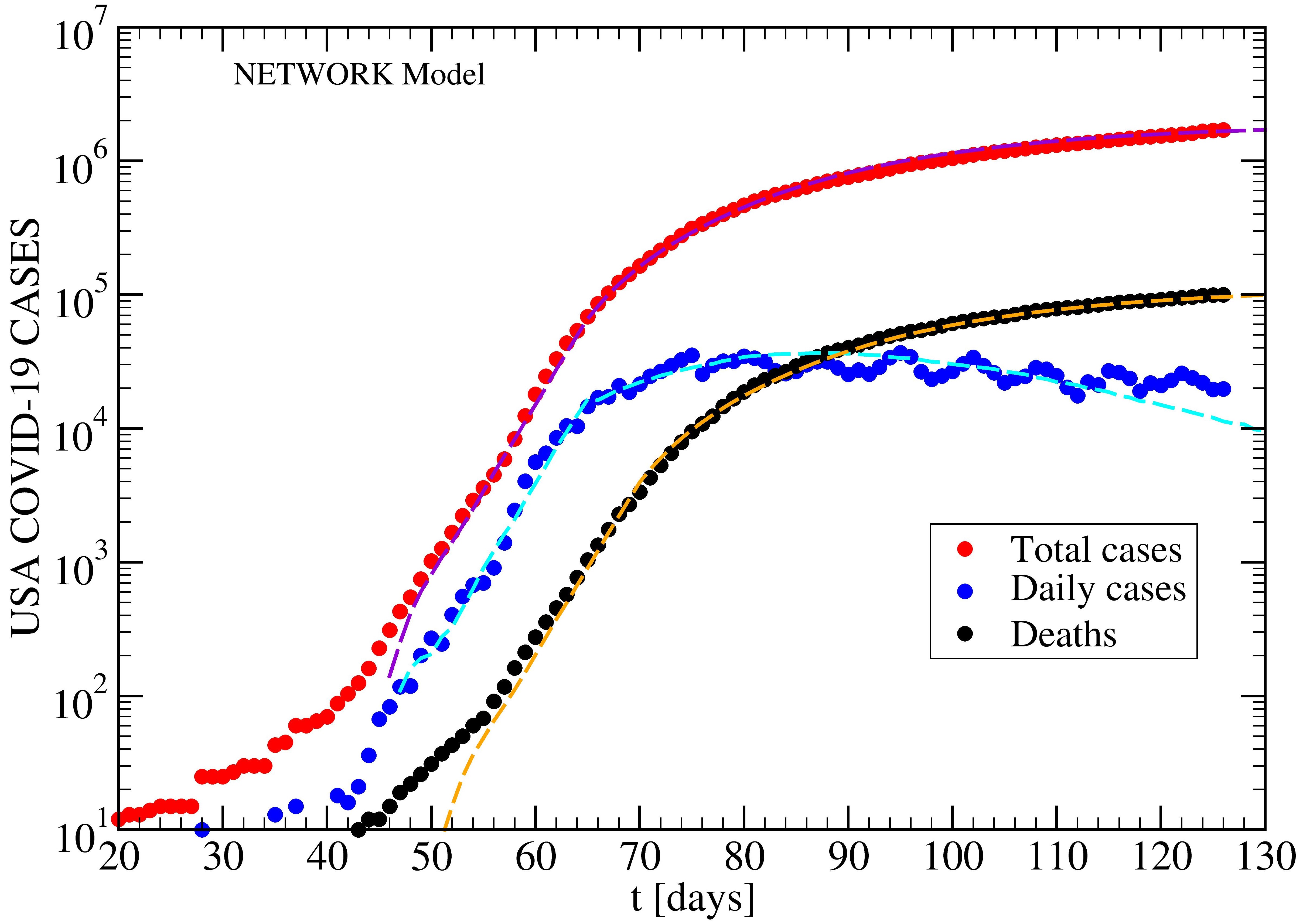}
\caption{Time evolution of COVID-19 in the USA. {\bf SIR model} (upper panel): $N=3\,10^6$, $f_0=1/2$, $\beta=0.65$,
             $\gamma=1/30$, $\bar R_0=9.75$; Lockdowns: $\tau_0=36$, $\beta_{\rm D}=0.45$, $q(t)=2.5-t/100$, 
             $d(t)=0.17 (0.3+1/(1+t/60))$. Time lag $t_{\rm Lag}=29$. 
             {\bf Network model} (lower panel): $L=400$, $N=90\, 10^6$, $\beta=0.65$, $\gamma =1/30$, $\bar R_0=9.75$,
             $\beta_{\rm L}=1/8$, $\left<k\right>=1.994$, $\left<\Delta k\right>=661/79900\cong0.008$ and 
             $\left<k_{\rm eff}\right>=2.002$; 
             Lockdowns: $\tau_0=20$, $\beta_{\rm D}=0.55$, $\beta_{\rm LD}=1/32$, $q=2.5$, $d=0.06$. Time lag $t_{\rm Lag}=45$ for 
             cases and $t_{\rm Lag}=21$ for deaths. Data up to May 25, 2020.}
\label{fig:USAcovidData}
\end{center}
\end{figure}

For the USA data (Fig.~\ref{fig:USAcovidData}), the lockdown regime can be described by the Ansatz, similar to Eq.~(\ref{eq:barbetatime}),
\begin{equation}             
\bar\beta(t) = \bar\beta_{\rm D} \left(\frac{\tau_0}{t}\right)^{q(t)}, \, t\ge \tau_0,  
\label{eq:barbetatimeUSA}
\end{equation}
where $\bar\beta_{\rm D}=(\beta_{\rm D}/\gamma) (1-f_0)$, with a time dependent exponent $q$ only for the SIR model.

Let us consider first the case of SIR (upper panel in Fig.~\ref{fig:USAcovidData}). For the latter, we use the decreasing function with time,
$q(t)=q-t/100$, with $q=2.5$. This feature was needed in order to reproduce the slowly decreasing behavior of the daily cases (blue circles in Fig.~\ref{fig:USAcovidData}). In addition, a rather complicated form for the factor $d(t)$ determining the time evolution of deaths, Deaths$=d(t)\times R(t)$, was required. We find that the form $d(t)=0.17[0.3+1/(t/60)]$ reproduces the curve of deaths quite well (black circles in Fig.~\ref{fig:USAcovidData}), but the results might be improved using more parameters. This remains to be understood. The whole fitting curves were shifted in time by an amount $t_{\rm Lag}=29$~days, that is, the initial data were actually discarded from the 
fits. 

\newpage

In the case of the network model (lower panel in Fig.~\ref{fig:USAcovidData}), the situation is simpler since all parameters can be
taken as constants, the values depending on the regime under consideration. Regarding lockdowns ($t\ge\tau_0$), the value $q=2.5$ works rather well, while we take a finite long-range transmission probability $\beta_{\rm LD}=1/32$, i.e. 4 times smaller than its value 
$\beta_{\rm L}=1/8$ ($t<\tau_0$), suggesting that indeed the lockdowns are not fully implemented and few additional infecteds are still
moving around. Also the number of deaths can be estimated from the actual recovereds using a single value $d=0.06$ (6\%). As well as 
in the case of SIR, also here we used times lags for the fits, i.e. $t_{\rm Lag}=45$ for the cases and $t_{\rm Lag}=21$ for deaths.

\section{Conclusions}
\label{sect:conclusions}

We have introduced a network model for infectious disease spreading in a population based on random graph and percolation theory concepts. The model is conveniently defined on a square lattice which permits a simple visualization of the four different categories
in the problem: Infecteds, susceptibles, recovereds and dormants. The first three groups form the core of the widely used SIR model, while
the fourth one is introduced here representing those individuals who are disconnected to some extent from the rest of the population. They
do not participate in the spreading phenomena but their presence acts as an effective slowing down of spreading by blocking an otherwise direct transmission between infecteds and susceptibles. In the language of percolation theory, the `connected' susceptibles
form finite clusters on the lattice which are separated from each other. To allow the spreading to overcome these `connection gaps' as the process evolves in time, we allow infecteds to reach any other susceptible site in the lattice, and infect it with a relatively lower probability than inside a finite cluster via NN contacts. We denote these new links, not present initially in the `lattice graph', dynamical links. This dynamical approach allows us to describe lockdown effects in terms of the slowing down or total lacking of the dynamical links.

We have assessed the performance of the network model by fitting COVID-19 USA data and compared the results with predictions using the SIR model. The network model works very well by just using constant parameters, while the SIR model requires more involved time dependent parameters to achieved similar fitting accuracy. We conclude that the present network model can become a valuable technique to complement the widely used SIR model.








\end{document}